\documentclass[12pt]{article}
\usepackage{fleqn}

\usepackage{graphicx}
\usepackage[figuresright]{rotating}

\newcommand{\AmS}{{\protect\the\textfont2
  A\kern-.1667em\lower.5ex\hbox{M}\kern-.125emS}}

\title{Radiative decays of light vector mesons. }
\author{
M.N.~Achasov, S.E.~Baru, K.I.~Beloborodov, \\ A.V.~Berdyugin,
A.V.~Bozhenok, A.D.~Bukin, D.A.~Bukin, \\ S.V.~Burdin, T.V.~Dimova,
S.I.~Dolinsky,  A.A.~Drozdetsky, \\ V.P.~Druzhinin, M.S.~Dubrovin,
I.A.~Gaponenko,
V.B.~Golubev, \\ V.N.~Ivanchenko, I.A.~Koop, A.A.~Korol, S.V.~Koshuba, \\
G.A.~Kukartsev, E.V.~Pakhtusova, A.A.~Polunin, V.M.~Popov, \\ A.A.~Salnikov,
S.I.~Serednyakov, V.V.~Shary, Yu.M.~Shatunov, \\ V.A.~Sidorov, Z.K.~Silagadze,
Yu.S.~Velikzhanin, \\ A.V.~Vasiljev, A.S.~Zakharov\\
{\em Budker Institute of Nuclear Physics, Novosibirsk 630090, Russia,}\\
{\em Novosibirsk State University, Novosibirsk 630090, Russia}\\
{ ~}\\
{talk given by S.~Burdin} \\ {at 8th International Conference on
   Hadron Spectroscopy (HADRON 99),}\\{ Beijing, China, 24-28 Aug 1999}
}
\date{}
\begin{document}
\maketitle
\begin{abstract}
The new data on $\rho,\omega,\phi$ radiative decays into 
$\pi^0\gamma,\eta\gamma,\eta'\gamma$ from SND experiment at VEPP-2M
$e^+e^-$ collider are presented.
\end{abstract}
\section{Introduction}
  Radiative decays of $\rho,\omega$ and $\phi$  mesons were studied in
  the SND experiment at the VEPP-2M $e^+e^-$ collider \cite{vepp2m}.
  Approximately $20\times 10^6$ $\phi$, $2\times 10^6$ $\rho$ and 
  $1\times 10^6$ $\omega$ mesons were collected. The results for
  $\phi$-meson decays were obtained mainly from the part of the data
  collected during 1996, when approximately $8\times 10^6$
   of $\phi$ mesons were produced with appropriate integrated
  luminosity 4.3~pb$^{-1}$.

  The SND detector \cite{SND} provides good possibility to study the
  radiative 
  decays of light vector mesons $e^+e^-\to \mathrm{V} \to \pi^0\gamma,
  \eta\gamma, \eta'\gamma$, where
  $\mathrm{V}=\rho,\omega,\phi$. In particular, large calorimeter
  solid angle, about 90\% of $4\pi$,  allows to select the
  multi-photon final states with high efficiency.

\section{Decays $\phi\to\eta\gamma,\pi^0\gamma$}

  The decay
  $\phi\to\eta\gamma$ was studied in the main final states of $\eta$:
$3\gamma,3\pi^0\gamma,\pi^+\pi^-\pi^0\gamma$, which cover about 94\%
  of all decays of $\eta$ meson.
The study of $3\gamma$ final state allows  to 
measure the probability of the decay $\phi\to\pi^0\gamma$ also. 

\subsection{$3\gamma$ final state \cite{Salnikov}.}

  In this final state two radiative decays of $\phi$ meson were studied: 
$e^{+}e^{-}\rightarrow \eta \gamma \rightarrow \gamma \gamma \gamma$
and
$e^{+}e^{-}\rightarrow \pi ^{0}\gamma \rightarrow \gamma \gamma \gamma$
with the main background coming from non-resonant QED three-quanta 
annihilation 
$e^{+}e^{-}\rightarrow \gamma \gamma \gamma \: (\mathrm{QED})$.


Preliminary selection included presence of three or
four reconstructed photons; cut on the total energy deposition in the
calorimeter: $0.7\cdot 2E_{beam}<E_{tot}<1.2\cdot 2E_{beam}$; cut on the
sum of the photon momenta: $\sum P_i<0.2E_{tot}/c$; cut on the minimal
energy of the photons at the level of 50~MeV. To suppress spurious
signals in the calorimeter, which appear mainly in the crystals
closest to the beam, additional restrictions were imposed on the
energies and angles of the reconstructed photons:
polar angle for the
  photons with energies 50--100~MeV was in the range \( 45^{\circ
    }<\theta <135^{\circ } \), while for the photons with the energies
  higher than 100~MeV it was in the range 
  \( 27^{\circ }<\theta <153^{\circ } \).

To distinguish between the 
 processes $\eta\gamma\to3\gamma$, $\pi^0\gamma\to3\gamma$ and 
$e^+e^-\to3\gamma\: (QED)$ the kinematic fit was used.
 About 18000 events were selected for the process
 $\eta\gamma\to3\gamma$ and about 1700 events for the process 
 $\pi^0\gamma\to3\gamma$ with corresponding efficiencies 44\% and
 14\%.

\subsection{$7\gamma$ final state \cite{3pi0gamma}}
In this final state  main background comes from the process
$\phi\to K_S K_L \to2\pi^0+X $. 
 The selection criteria included presence of 6--8 reconstructed
 photons, cuts on the total energy deposition and the momentum
 balance. The most energetic photon in the event is the recoil photon
with energy about 360~MeV. Therefore  the restriction on the energy of
the most energetic photon was imposed.

  The selection efficiency of this final state was about 32\%. 
  Approximately 10000 events were
  selected with the background lower than 1\%.
 
\subsection{$\pi^+\pi^-\pi^0\gamma$ final state \cite{3pi}} 
Main background for this final state comes
  from the decay $\phi\to\pi^+\pi^-\pi^0$ with spurious hits in calorimeter.

At first the events with 2 charged particles and 3 or more photons
were selected. Then the cuts on distances between charged tracks and
beam axis and on space angle between charged tracks (to reject the
events of the 
process $\phi\to K_S K_L \to \pi^+\pi^- + X$)
 were applied.
To suppress the background the kinematic fit was used.

  The selection efficiency of this final state was about 18\%.
Approximately $20\times 10^{6}$ events of $\phi$-meson 
decays were processed. 

\subsection{Analysis}

For the description of the cross section of processes \(
e^{+}e^{-}\rightarrow P\gamma \), where \( P \) is a pseudo-scalar
meson, the 
following dependence was used:
\begin{equation}
\label{eq:cross-section-general}
\sigma (s)  =  \frac{F(s)}{s^{3/2}}\left|\sum _{V=\rho,\omega,\phi}
\sqrt{\sigma _{VP\gamma}\frac{m_{V}^{3}}{F(m_{V}^{2})}}
\frac{m_V\Gamma_V e^{\mathrm{i}\varphi
_V}}{m_V^2-s-\mathrm{i}\sqrt{s}\Gamma _V(s)}
\right| ^{2}, 
\end{equation}
where $\sigma _{VP\gamma }=12\pi B (V \to e^+e^-) B (V\to P\gamma)/m_{V}^{2}$
is the cross section of
the process $e^+e^-\to V\to P\gamma$ at the maximum of vector resonance $V$.
$F(s)=[(s-m_{P}^2)/2\sqrt{s}]^3$ 
is the phase space factor for the process $e^+e^-\to P\gamma$.
The relative phases of vector mesons were taken to be $\varphi _{\rho}=\varphi
_{\omega}=0$, $\varphi _{\phi}=180^\circ$ for $\eta\gamma$ decay,
$\varphi _{\phi}=(158\pm11)^\circ$ for $\pi^0\gamma$ decay.

 The fit gave the following results for the decay 
 $\phi\to\eta\gamma$ in the different final states: 
\begin{eqnarray}
\label{eq:br:eta_gamma:3gamma}
3\gamma & : & \mathrm{BR}(\phi\to\eta\gamma)=(1.338\pm 0.012\pm 0.052)\% \\
\label{eq:br:eta_gamma:7gamma}
7\gamma & : & \mathrm{BR}(\phi\to\eta\gamma)=(1.296\pm 0.024\pm 0.057)\% \\
\label{eq:br:eta_gamma:pippimpi0}
\pi^+\pi^-\pi^0 &:&\mathrm{BR}(\phi\to\eta\gamma)=(1.259\pm 0.030\pm 0.059)\% 
\end{eqnarray}

  Main sources of systematic errors 
  were luminosity measurement (2.5\%), error in
  $\mathrm{BR}(\phi\to e^+e^-)$ (3\%),
  MC efficiency determination (1--2\%),
  the error in the branching ratios of the decays of $\eta$ meson  
  (1--2\%) and model dependence (1.5\%).

 If we combine the branching ratios (\ref{eq:br:eta_gamma:3gamma}), 
 (\ref{eq:br:eta_gamma:7gamma}) and (\ref{eq:br:eta_gamma:pippimpi0}) 
 some systematic errors will cancel and 
we will obtain $\mathrm{BR}(\phi\to\eta\gamma)=(1.304\pm 0.049)\%$. 
This result agrees with the world average
$\mathrm{BR}(\phi\to\eta\gamma)=(1.26\pm 0.06)\%$ \cite{PDG} and has smaller
error. The systematic error comes mainly from the error in the branching
ratio $\mathrm{BR}(\phi\to e^+e^-)$.

For the decay $\phi\to\pi^0\gamma$ the  following result  was obtained:
$
\mathrm{BR}(\phi\to\pi^0\gamma)=(1.226\pm
0.036_{-0.089}^{+0.096})\times 10^{-3} $.
  The uncertainty in phase
  $\varphi_\phi$ for the process $\phi\to\pi^0\gamma$ gave about
  6\% systematic error for the decay $\phi\to\pi^0\gamma$.
\begin{figure}[hbt]
\begin{minipage}[t]{75mm}
\includegraphics[width=70mm]{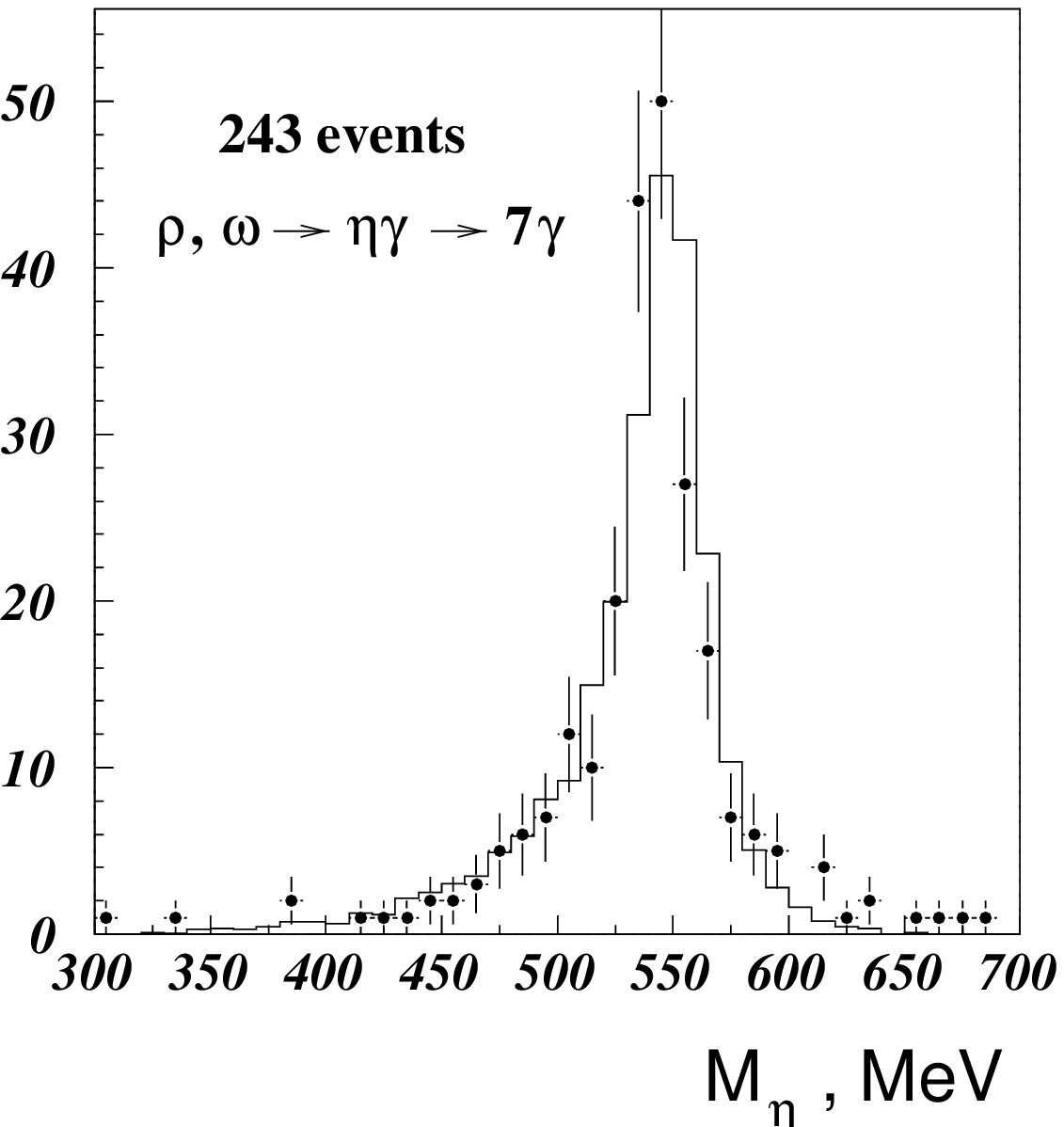}
\caption{ Recoil mass of the most energetic photon in an event of the
process $e^+e^-\to\eta\gamma\to7\gamma$.
 The histogram --- MC; the points --- experiment.}
\label{fig:proc:receta}
\end{minipage}
\hspace{\fill}
\begin{minipage}[t]{75mm}
\includegraphics[width=74mm]{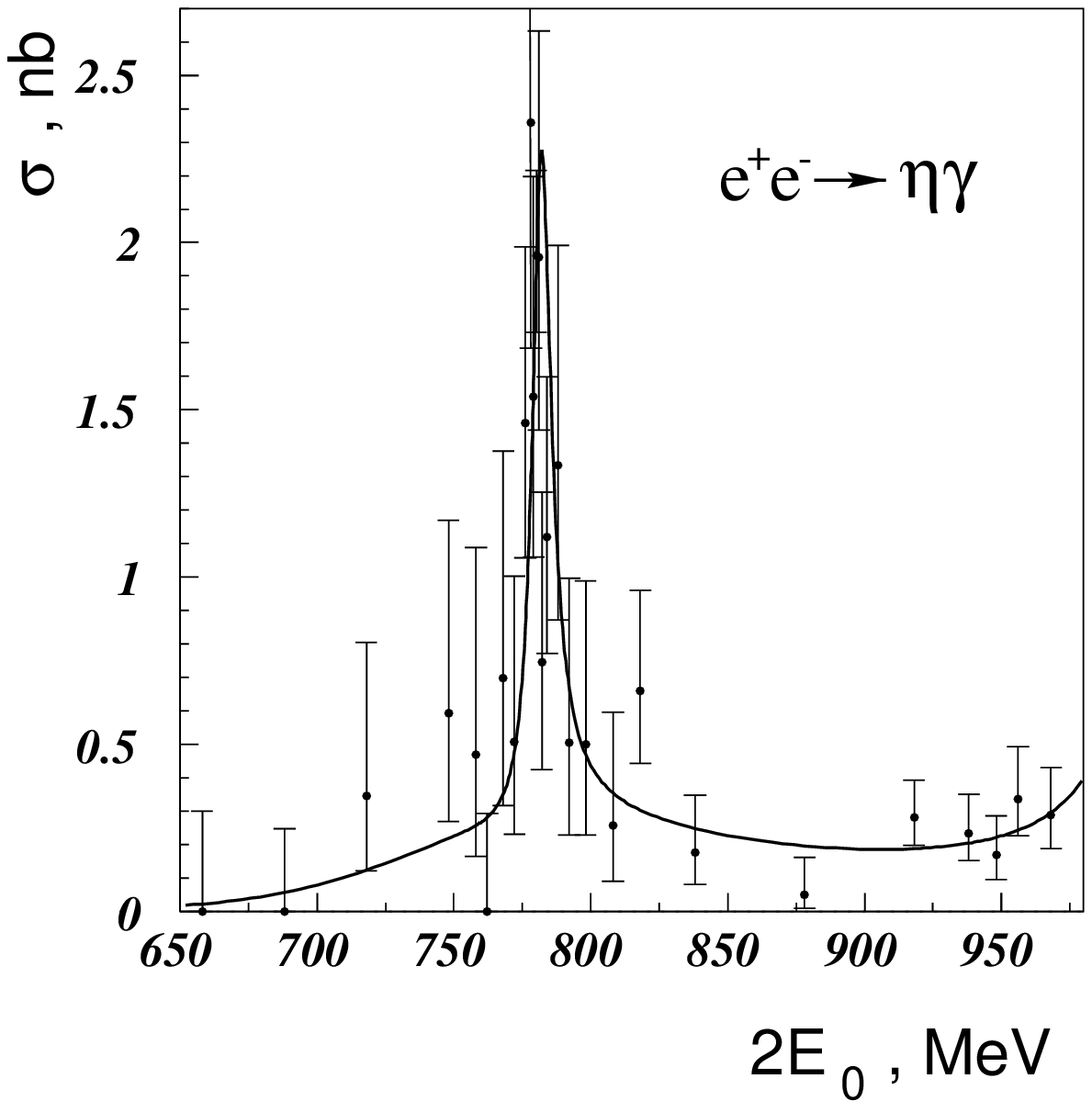}
\caption{ Born cross section of the process $e^+e^-\to\eta\gamma$.}
\label{fig:cros:etagam}
\end{minipage}
\end{figure}

\section{Decay $\phi\to\eta'\gamma$ \cite{etaprgamma}} 

The first observation of this decay was done at
VEPP-2M in CMD-2 experiment~\cite{Akhmetshin97}. The measurement of
this decay at SND was performed with $\eta'$ decaying into
$\pi^+\pi^-\eta$ and $\eta$ into two $\gamma$'s. 
The background for this final state comes from the processes
$e^+e^-\to\eta\gamma\to\pi^+\pi^-\pi^0\gamma$,
$e^+e^-\to\pi^+\pi^-\pi^0$ and
$e^+e^-\to\omega\pi^0\to\pi^+\pi^-\pi^0\pi^0$. For the analysis 
events with two charged tracks and three photons were selected. To
suppress the background a complex selection algorithm was developed
based on the kinematics of all these processes. It was described in
details in the ref.~\cite{etaprgamma}. Due to the strict cuts
 the selection efficiency of the final state
$\pi^+\pi^-3\gamma$ for the events of the process under study was 5.5\%. 

There were found
 $5.2^{+2.6}_{-2.2}$ events, which correspond to the 
branching ratio
$\mathrm{BR} (\phi\to\eta'\gamma) = (6.7^{+3.4}_{-2.9}) \times 10^{-5}$.
The systematic error, not included in the above errors, is about 15\%
and determined mainly by the error in the efficiency estimation.

\section{Preliminary results on decays
 $\rho,\omega\to\eta\gamma$} 
 To study the decays  $\rho,\omega\to\eta\gamma$ the final
 state $\eta\to3\pi^0\to6\gamma$ was chosen because of physical background
  in the energy region of $\rho$  meson is absent for this final state.
  The analysis of these decays is
  similar to the analysis of the decay
  $\phi\to\eta\gamma\to3\pi^0\gamma\to7\gamma$. The spectrum of  recoil
  mass of the most energetic photon in the event of the process
  under study is shown in
  Fig.~\ref{fig:proc:receta}. The Born cross section of the process
  $e^+e^-\to\eta\gamma$ in the energy region of $\rho$ meson is
  presented in Fig.~\ref{fig:cros:etagam}. The fit was done by the
  formula (\ref{eq:cross-section-general}) with fixed phases
$\varphi _{\rho}=\varphi_{\omega}=0$, $\varphi _{\phi}=180^\circ$.
  The resulting branching ratios are following:
$\mathrm{BR}(\omega\to\eta\gamma)=(5.9\pm 1.0)\times 10^{-4}$,
$\mathrm{BR}(\rho\to\eta\gamma)=(2.0\pm 0.4)\times 10^{-4}$. 
  These results agree with the table
  values $\mathrm{BR}(\omega\to\eta\gamma)=(6.5\pm 1.0)\times 10^{-4}$ and
  $\mathrm{BR}(\rho\to\eta\gamma)=(2.4^{+0.8}_{-0.9})\times 10^{-4}$
  \cite{PDG}. The result on the decay $\rho\to\eta\gamma$ has smaller
  error.

\section{Summary on the results}
The results presented in this work are summarized in the table: \\
\newcommand{\m}{\hphantom{$-$}}
\newcommand{\cc}[1]{\multicolumn{1}{c}{#1}}
\renewcommand{\tabcolsep}{2pc} 
\renewcommand{\arraystretch}{1.2} 
\begin{tabular}{|c|c|c|}
\hline
Decay           & BR(SND) & BR(PDG) \cite{PDG} \\
\hline
$\phi\to\eta\gamma$       & $(1.304\pm0.049)\%$ 
& $(1.26\pm0.06)\%$ \\
$\phi\to\pi^0\gamma$      &
$(1.226\pm0.036^{+0.096}_{-0.089})\times10^{-3}$ 
& $(1.31\pm0.13)\times10^{-3}$  \\
$\phi\to\eta'\gamma$      & $(6.7^{+3.4}_{-2.9})\times10^{-5}$
& $(12^{+7}_{-5})\times10^{-5}$ \\
$\omega\to\eta\gamma$     & $(5.9\pm1.0)\times10^{-4}$
& $(6.5\pm1.0)\times10^{-4}$ \\
$\rho\to\eta\gamma$       & $(2.0\pm0.4)\times10^{-4}$
& $(2.4^{+0.8}_{-0.9})\times10^{-4}$ \\
\hline
\end{tabular} 

\section{Acknowledgement}
 The work is partially supported by RFBR (Grants No 96-15-96327, 99-02-17155,
 99-02-16815, 99-02-16813) and STP ``Integration'' (Grant No 274).

\end{document}